\documentclass[12pt]{article}

\usepackage[totalheight = 23cm, totalwidth = 17cm]{geometry}
\usepackage{amssymb,amsmath,amsfonts,amsbsy,graphicx}

\newcommand{\mpl}{m_{\rm Pl}}
\newcommand{\fnl}{f_{\rm NL}}

\newcommand{\calN}{{\cal N}}

\begin{document}

\begin{titlepage}

\begin{center}

\vspace*{-10ex}
\hspace*{\fill}
CERN-PH-TH/2011-012, YITP-11-9

\vskip 1.5cm

\Huge{A covariant approach \\ to general field space metric \\ in multi-field inflation}

\vskip 1cm

\large{
Jinn-Ouk Gong$^{*}$\footnote{jinn-ouk.gong@cern.ch}
\hspace{0.2cm}\mbox{and}\hspace{0.2cm}
Takahiro Tanaka$^{\dag}$\footnote{tanaka@yukawa.kyoto-u.ac.jp}
\\
\vspace{0.5cm}
{\em
${}^*$ Theory Division, CERN
\\
CH-1211 Gen\`eve 23, Switzerland
\\
\vspace{0.2cm}
${}^\dag$ Yukawa Institute for Theoretical Physics, Kyoto University
\\
Kyoto 606-8502, Japan}
}

\vskip 0.5cm

\today

\vskip 1.2cm

\end{center}

\begin{abstract}

We present a covariant formalism for general multi-field system which enables us to
obtain higher order action of cosmological perturbations easily and
systematically. The effects of the field space geometry,
described by the Riemann curvature tensor of the field space, are naturally
incorporated. We explicitly calculate up to the cubic order action which
is necessary to estimate non-Gaussianity and present those geometric
terms which have not yet known before.

\end{abstract}

\end{titlepage}

\setcounter{page}{0}
\newpage
\setcounter{page}{1}

\section{Introduction}
\label{sec:intro}

Inflation~\cite{inflation} is currently the leading candidate to lay down the necessary initial conditions for the successful hot big bang evolution of the universe~\cite{book}. The most recent observations from the cosmic microwave background (CMB) are consistent with the predictions of the inflationary paradigm~\cite{Komatsu:2010fb}: the universe is homogeneous and isotropic with vanishing spatial curvature, and the primordial scalar perturbation is dominantly adiabatic and follows almost perfect Gaussian statistics with a nearly scale invariant power spectrum. Thus, any small deviation from these predictions would provide crucial information for us to distinguish different models of inflation. Especially, the non-linearities in the primordial perturbation have received an extensive interest nowadays in the light of upcoming precise cosmological observations. For example, while the current bound on the non-linear parameter $\fnl$~\cite{Komatsu:2001rj} is constrained to be $|\fnl| \lesssim \mathcal{O}(100)$ from the Wilkinson Microwave Anisotropy Probe observation on the CMB~\cite{Komatsu:2010fb}, the Planck satellite can probe with better precision to detect $|\fnl| = \mathcal{O}(5)$~\cite{:2006uk}. The sensitivity may be even further improved from the observations on large scale structure~\cite{LSSfNL}.

The absence of the relevant scalar field which can support inflation in the standard model (SM) of particle physics\footnote{
It was recently suggested that the SM Higgs field can play the role of the inflaton provided that it is non-minimally coupled to gravity~\cite{Bezrukov:2007ep}. However the unitarity of the simplest Higgs inflation appears to be controversial. See e.g. Refs.~\cite{unitX} and \cite{unitO} and references therein for different points of view on this issue.
} demands that inflation be described in the context of the theories beyond the SM. Typically there are plenty of scalar fields which can contribute to the inflationary dynamics~\cite{Lyth:1998xn}. Further, in multi-field system we can obtain interesting observational signatures which deviate from the predictions of the single field models of inflation and can be detected in near future, such as isocurvature perturbation~\cite{Choi:2008et} or non-Gaussianity~\cite{nGreviews}. Thus we have both theoretical and phenomenological motivations to develop a complete formulation of general multi-field inflation.

An important point in multi-field system is that in the field space which generally has non-trivial field space metric, the scalar fields play the role of the coordinate. Naturally, as we do in general relativity, it is preferable to formulate the dynamics in the field space in the coordinate independent manner. That is, we need a covariant formulation of multi-field inflation which allows us to describe the inflationary dynamics with arbitrary field space. However, most studies on multi-field inflation, especially regarding non-linear perturbations, are based on trivial field space~\cite{Seery:2005gb} or non-covariant description~\cite{Langlois:2008wt,Langlois:2008qf,Arroja:2008yy}. The existing studies with covariant approach to general field space metric are mostly on linear perturbation theory~\cite{linearcov}.

In this note, we develop a fully covariant formulation of non-linear perturbations in general multi-field inflation. Along with the covariance for general field space, it allows us to obtain arbitrary higher order action of cosmological perturbations easily and systematically. We consider the matter Lagrangian which is a generic function of the field space metric $G_{IJ}$ with $I$ and $J$ being generic field space indices, kinetic function $\partial^\mu\phi^I\partial_\mu\phi^J$ and the fields~\cite{kinflation}. This form includes not only the matter Lagrangian with the standard canonical kinetic term but also more generic ones motivated from high energy theories, such as the Dirac-Born-Infeld (DBI) type~\cite{DBI}.

This note is outlined as follows. In Section~\ref{sec:mapping} we set up the geodesic equation to describe the field fluctuation around the background trajectory. In Section~\ref{sec:matterL}, we consider pure matter Lagrangian and present a covariant formulation to describe the field fluctuations up to arbitrary order. The extension to include gravity follows in Section~\ref{sec:gravity} and we explicitly compute the perturbed action up to cubic order. We also discuss the genuine multi-field effects briefly. We conclude in Section~\ref{sec:conclusions}. Technical details to compare with the previously known non-covariant description are presented in the Appendix.

\section{Issue of mapping}
\label{sec:mapping}

\begin{figure}[h]
 \centering
 \includegraphics[width=10cm,angle=90]{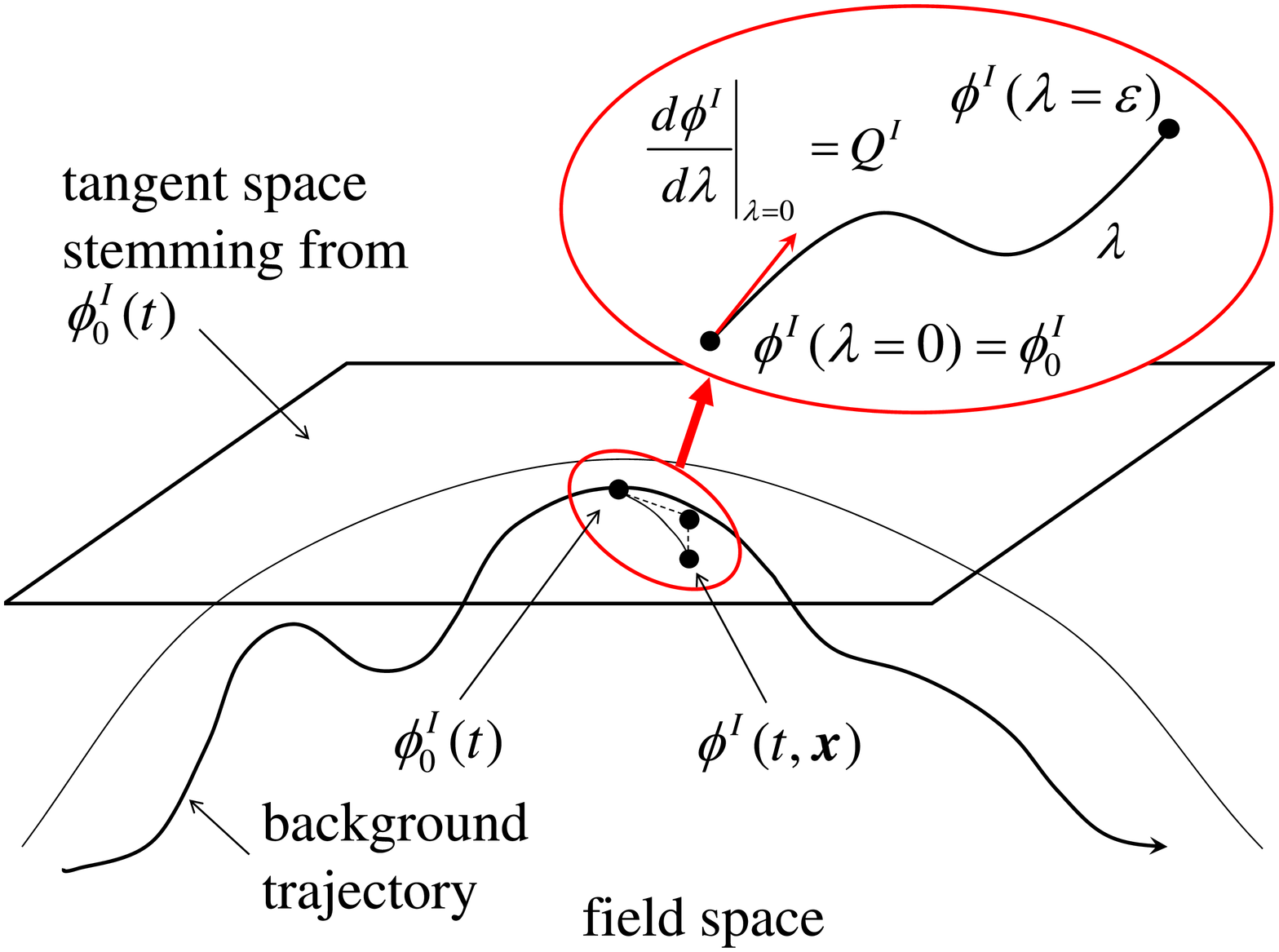}
 \caption{A schematic figure showing a physical field $\phi^I$ in the field space around the background trajectory $\phi_0^I(t)$. The geodesic connecting $\phi^I$ and $\phi_0^I$ is parametrized $\lambda$, which runs from 0 to $\epsilon$.}
 \label{fig:field_space}
\end{figure}

To begin with, first let us consider how to describe the physical field
fluctuation $\delta\phi^I$ in the field space in a covariant manner.
We can think of the background
field trajectory parametrized by a single parameter, usually taken as
the cosmic time $t$: $\phi_0^I = \phi_0^I(t)$. The real physical field
in a fixed gauge $\phi^I$ incorporates quantum fluctuations
$\delta\phi^I$
around this background trajectory.
However, the fluctuations $\delta\phi^I$ are coordinate dependent,
and hence they are not covariant.
These two points, $\phi_0^I(t)$ and $\phi^I$, can
be connected by a unique geodesic with respect to
the field space metric $G_{IJ}$ as long as their separation
is sufficiently small. This geodesic can be specified by
the initial point $\phi^I_0$ and its initial velocity,
which we denote by $Q^I$.
This situation is depicted in Figure~\ref{fig:field_space}.
Hence, the issue here is the ``mapping'' beyond linear order
between the finite displacement $\delta\phi^I\equiv\phi^I-\phi^I_0$
and a vector $Q^I$ living in the tangent space at $\phi^I_0$.
Let us parametrize the geodesic trajectory  in the field space by
$\lambda$, which runs from 0 to $\epsilon>0$: $\lambda=0$ and
$\lambda=\epsilon$ correspond to
$\phi_0^I$ and $\phi^I$, respectively. Here $\epsilon$ is a parameter
introduced to count the order of perturbation
just for a bookkeeping purpose, and hence
it is set to unity at the end of calculation.

Denoting the covariant differentiation in $\lambda$-direction by
$D_\lambda \equiv D/d\lambda$,
the geodesic equation for $\phi^I(\lambda)$ is written as
\begin{equation}\label{geodesic}
D_\lambda^2\phi^I = \frac{d^2\phi^I}{d\lambda^2} + \Gamma^I_{JK}\frac{d\phi^J}{d\lambda}\frac{d\phi^K}{d\lambda} = 0 \, ,
\end{equation}
and the initial conditions are
\begin{align}
\left. \phi^I \right|_{\lambda=0} = & \phi_0^I \, ,
\\
\left. D_\lambda\phi^I \right|_{\lambda=0} = & \left. \frac{d\phi^I}{d\lambda} \right|_{\lambda=0} = Q^I \, .
\end{align}
Now,
we expand $\phi^I(\lambda=\epsilon)$ as a power series with respect to
$\epsilon$ from $\lambda=0$ as
\begin{equation}\label{mapping}
\phi^I(\lambda=\epsilon) = \left. \phi^I \right|_{\lambda=0} + \left. \frac{d\phi^I}{d\lambda} \right|_{\lambda=0} \epsilon + \left. \frac{1}{2!}\frac{d^2\phi^I}{d\lambda^2} \right|_{\lambda=0} \epsilon^2 + \left. \frac{1}{3!}\frac{d^3\phi^I}{d\lambda^3} \right|_{\lambda=0} \epsilon^3 + \cdots \, .
\end{equation}
Note that the derivatives with respect to $\lambda$ here are {\em not}
covariant ones. Thus, we can trade quadratic and higher derivatives with
single derivatives
by means of the geodesic equation (\ref{geodesic}).
Namely, we can replace a quadratic derivative with
\begin{equation}\label{geodesic3}
\frac{d^2\phi^I}{d\lambda^2} = -\Gamma^I_{JK}\frac{d\phi^J}{d\lambda}\frac{d\phi^K}{d\lambda} \, ,
\end{equation}
and the third order derivative with
\begin{equation}
\frac{d^3\phi^I}{d\lambda^3} = \left( \Gamma^I_{LM}\Gamma^M_{JK} - \Gamma^I_{JK;L} \right) \frac{d\phi^L}{d\lambda}\frac{d\phi^J}{d\lambda}\frac{d\phi^K}{d\lambda} \, ,
\end{equation}
and so on. Thus, we can write (\ref{mapping}) as
\begin{equation}
\phi^I(\lambda=\epsilon) = \phi^I_0 + Q^I\epsilon - \frac{1}{2}\Gamma^I_{JK}Q^JQ^K \epsilon^2 + \frac{1}{6} \left( \Gamma^I_{LM}\Gamma^M_{JK} - \Gamma^I_{JK;L} \right) Q^JQ^KQ^L \epsilon^3 + \cdots \, ,
\end{equation}
which we can continue up to arbitrary non-linear order.
In the end, setting $\epsilon = 1$, we obtain
\begin{equation}\label{mapping2}
\phi^I-\phi_0^I \equiv \delta\phi^I = Q^I - \frac{1}{2}\Gamma^I_{JK}Q^JQ^K + \frac{1}{6} \left( \Gamma^I_{LM}\Gamma^M_{JK} - \Gamma^I_{JK;L} \right) Q^JQ^KQ^L + \cdots \, .
\end{equation}
If we truncate (\ref{mapping2}) at linear order, we can identify
$\delta\phi^I$ and $Q^I$. Then, we do not have to pay attention to the
difference between them. However, when we consider non-linear
perturbations, we have to distinguish them clearly. Only when we write
the equations in terms of $Q^I$, they can be expressed in a covariant
manner.

\section{General matter Lagrangian}
\label{sec:matterL}

Now, let us consider the general effective matter Lagrangian $P$, which is a function of the field space metric $G_{IJ}$, kinetic function $X^{IJ} = -g^{\mu\nu}\partial_\mu\phi^I\partial_\nu\phi^J/2$ and $\phi^I$, i.e.
\begin{equation}
P = P(G_{IJ},X^{IJ},\phi^I) \, .
\end{equation}
Note that we do not restrict the kinetic function to a function of $X = G_{IJ}X^{IJ}$,
i.e. all the indices contracted with the metric. This is because we may
have a term like $G_{IK}G_{JL}X^{IJ}X^{KL}$, as is typical in the
multi-field DBI inflation.
Also we do not consider higher derivative terms
such as $\Box\phi^I$, which usually leads to ghost except for
some special combinations such as Galileon~\cite{Nicolis:2008in}.
In this section, we treat the spacetime metric $g_{\mu\nu}$ as a given background,
but it is not necessarily spatially homogeneous.
Inclusion of the metric perturbation will be
discussed in the succeeding section.

We consider that the fields $\phi^I$ contained in $P$ are all functions
of $\lambda$, a parameter along the geodesic in the field space.
Then, $P$ as a whole is a function of $\lambda$ and is a scalar with respect to the field space
indices. We expand $P$ in terms of the parameter $\lambda$, and set
$\lambda$ to $\epsilon$, to obtain
\begin{equation}\label{P_expansion}
P = P|_{\lambda=0} + \left. D_\lambda P \right|_{\lambda=0} \epsilon + \frac{1}{2!} \left. D_\lambda^2P \right|_{\lambda=0} \epsilon^2 + \frac{1}{3!} \left. D_\lambda^3P \right|_{\lambda=0} \epsilon^3 + \cdots \, ,
\end{equation}
where we have used the fact that an ordinary derivative of a field space
scalar is
identical to a covariant one.
First let us consider the linear variation, $D_\lambda P$.
At this stage, we find
\begin{equation}\label{P_expansion_linear}
D_\lambda P =
\frac{\partial P}{\partial X^{IJ}} D_\lambda X^{IJ} +
\frac{\partial P}{\partial \phi^I}
D_\lambda \phi^I~,
\end{equation}
without any subtle issue.
Here we have used $D_\lambda G_{IJ} = 0$ that follows from the definition of
the covariant differentiation,
and we have also assumed that a derivative of $P$ with respect to
$X^{IJ}$ is automatically symmetrized, i.e.
\begin{equation}
\frac{\partial P}{\partial X^{IJ}} \to \frac{1}{2} \left(
   \frac{\partial P}{\partial X^{IJ}} + \frac{\partial P}{\partial
   X^{JI}} \right)
  \equiv P_{\langle IJ \rangle} \, .
\end{equation}

However, from the quadratic variation, we find that our notation becomes
a little uncomfortable. Explicitly, we can write
\begin{equation}\label{P_expansion_quadratic}
D_\lambda^2P = D_\lambda^2X^{IJ}P_{\langle{IJ}\rangle} + D_\lambda X^{IJ} \left( D_\lambda P_{\langle{IJ}\rangle} \right) + D_\lambda^2\phi^IP_{,I} + D_\lambda\phi^I \left( D_\lambda P_{,I} \right) \, ,
\end{equation}
where the third term vanishes due to the geodesic equation of
$\phi^I$. The difficulty is in the second and the last terms: how to write the
covariant derivatives of the derivative of $P$?
In fact, we can easily come to know
that the differentiation of $P$ with respect to $X^{IJ}$ should be understood
as an ordinary one because $X^{IJ}$ is not
a coordinate in the field space but a tensor living in the tangent
space. On the other hand, the differentiation with respect to $\phi^I$
should be understood as a covariant one because $\phi^I$ is a
coordinate of the field space:
any differentiation in the field space necessarily incorporates parallel
transport.

While the above considerations are legitimate, it is very uncomfortable
to have covariant and ordinary differentiations mixed.
Moreover, covariant and ordinary differentiations do not commute.
We have $P_{\langle{IJ}\rangle;K}$ coming from the second term of
(\ref{P_expansion_quadratic}) and
$P_{;K \langle{IJ}\rangle}$ coming from the last term,
but they are not the same.
Explicitly,
$P_{\langle{IJ}\rangle;K}
=P_{;K\langle{IJ}\rangle}
-\Gamma^L_{\,IK}P_{\langle LJ\rangle}-\Gamma^L_{\,JK}P_{\langle
IL\rangle}$.
Therefore if we rewrite one expression with the other, the result
contains the Christoffel symbols and is not manifestly covariant.

To avoid this mess, we consider an alternative description.
We assume that $P$
depends on $\phi^I$ only through field space tensors such as
${f}^{J_1\cdots J_{n_a}}_a(\phi^I)$,
where the subscript $a$ is introduced to discriminate
different kinds of such tensors.
One most important example is the potential $V(\phi^I)$, which is a
field space scalar.
Here we are assuming that there is no spacetime derivatives of fields
in ${f}^{J_1\cdots J_{n_a}}_a(\phi^I)$.
With this, first let us consider a
single derivative. From
\begin{equation}
P = P \left[ G_{IJ},X^{IJ},f^{J_1\cdots J_{n_a}}_a(\phi^I) \right] \, ,
\end{equation}
a single derivative with respect to $\lambda$ is easily calculated as
\begin{align}
D_\lambda P = & D_\lambda G_{IJ}\frac{\partial P}{\partial G_{IJ}} +
D_\lambda X^{IJ}\frac{\partial P}{\partial X^{IJ}} +
\sum_a D_\lambda f_a^{J_1\cdots J_{n_a}}\frac{\partial P}{\partial f^{J_1\cdots J_{n_a}}_a}
\cr
= & D_\lambda X^{IJ}P_{\langle{IJ}\rangle} + \sum_a
f_a^{J_1\cdots J_{n_a}}{}_{;I} D_\lambda\phi^I P_{\{J_1\cdots J_{n_a}\}_a}
\, ,
\end{align}
where we have defined $P_{\{J_1\cdots J_{n_a}\}_a} \equiv {\partial P}/{\partial
f^{J_1\cdots J_{n_a}}_a}$.
Now the differentiations of $P$ are all ordinary ones, and those of
$f^{J_1\cdots J_{n_a}}_a$ are all covariant ones.
In this way, we can straightforwardly write up to cubic order expansion
of the general matter Lagrangian $P$ with respect to $\lambda$ as
\begin{align}
P = & P|_{\lambda=0}
+ P_{\langle{IJ}\rangle} \delta X^{IJ} + P_a \delta f_a
+ {1\over 2!}P_{\langle{IJ}\rangle\langle{KL}\rangle} \delta X^{IJ}\delta
 X^{KL}
+ P_{\langle{IJ}\rangle{a}} \delta X^{IJ}\delta f_a
+{1\over 2!}P_{ab} \delta f_a \delta f_b
\nonumber\\
&
+ {1\over 3!}P_{\langle{IJ}\rangle\langle{KL}\rangle\langle{MN}\rangle}
   \delta X^{IJ}\delta X^{KL} \delta X^{MN}
+ {1\over 2!}P_{\langle{IJ}\rangle \langle{KL}\rangle {a}} \delta
 X^{IJ}\delta X^{KL} \delta f_a
\nonumber \\
&
 + {1\over 2!}P_{\langle{IJ}\rangle {ab}} \delta X^{IJ} \delta f_a  \delta f_b
+{1\over 3!}P_{abc} \delta f_a \delta f_b \delta f_c+\cdots,
\label{general_matter}
\end{align}
where we have assumed that the field space tensors $f^{J_1\cdots J_{n_a}}_a$
are all scalars for simplicity, and introduced the following notations
\begin{align}
 P_a \equiv &{\partial P\over \partial f_a} \, ,
 \\
 \delta X^{IJ} \equiv &\sum_{n=1}^\infty{\epsilon^n \over n!}D_\lambda^n X^{IJ}|_{\lambda=0} \, ,
 \\
 \delta f_a \equiv &\sum_{n=1}^\infty{\epsilon^n\over n!}D_\lambda^n f_a|_{\lambda=0} \, .
\end{align}
With the aid of the geodesic equation (\ref{geodesic}),
it is trivial to find that the derivatives of $f_a$ are given by
\begin{align}
\left. D_\lambda f_a \right|_{\lambda=0} = & f_{a;I}Q^I \, ,
\\
\left. D_\lambda^2f_a \right|_{\lambda=0} = & f_{a;IJ}Q^IQ^J \, ,
\\
\left. D_\lambda^3f_a \right|_{\lambda=0} = &
 f_{a;IJK}Q^IQ^JQ^K \, .
\end{align}
Obtaining the derivatives of $X^{IJ}$ with respect to $\lambda$
needs some manipulation.
First, we should understand that $\partial_\mu \phi^I$
is a vector living in
the tangent space. Hence, the covariant differentiation of $\partial_\mu
\phi^I$ is given by
\begin{equation}
 {D_\lambda} \partial_\mu\phi^I
  =\partial_\mu{d\phi^I\over d\lambda}+\Gamma^I_{~JK} \partial_\mu
  \phi^J {d\phi^K\over d\lambda}
  \equiv D_\mu {d\phi^I\over d\lambda} \, .
\end{equation}
When we recursively act the covariant differentiation $D_\lambda$, we
need the commutator between $D_\mu$ and $D_\lambda$. The necessary commutation
relation can be derived in the same manner as in the derivation of the geodesic
deviation equation, e.g. for an arbitrary vector $V^I$,
\begin{equation}
[D_\lambda,D_\mu] V^I=R^I_{~JKL} V^J {d\phi^K\over d\lambda}
 \partial_\mu \phi^L \, .
\end{equation}
Then, we obtain
\begin{align}
\label{expansionX1}
\left. D_\lambda X^{IJ} \right|_{\lambda=0} = & -g^{\mu\nu}
 D_\mu Q^{(I}\partial_\mu \phi_0^{J)} \, ,
\\
\label{expansionX2}
\left. D_\lambda^2X^{IJ} \right|_{\lambda=0} = &
-g^{\mu\nu}\left[
  R^{(I}{}_{KLM}\partial_\mu
 \phi_0^{J)}\partial_\nu\phi_0^MQ^KQ^L
  +D_\mu Q^I D_\nu Q^J\right] \, ,
\\
\left. D_\lambda^3X^{IJ} \right|_{\lambda=0} = &
 -g^{\mu\nu}\Bigl[
  R^{(I}{}_{KLM;N}\partial_\mu \phi_0^{J)}\partial_\nu \phi_0^MQ^NQ^KQ^L +
 R^{(I}{}_{KLM}\partial_\mu \phi_0^{J)}Q^KQ^LD_\nu Q^M
\cr &\hspace{1.2cm}+
 3R^{(I}{}_{KLM} D_\nu Q^{J)} \partial_\mu \phi_0^MQ^KQ^L
\Bigr] \, ,
\label{expansionX3}
\end{align}
where parentheses over the indices denote symmetrization. Here
we have written down explicitly how the inverse metric $g^{\mu\nu}$ is contained
in the expressions for the later convenience when we consider metric perturbations.

\section{Gravity}
\label{sec:gravity}

\subsection{General arguments}

Until now, we have only considered matter Lagrangian and treated the
metric as a given background. But to describe real physics we must take
into account the dynamics of gravitational degrees of freedom:
additional 4 scalar, 4 vector, and 2 tensor degrees of freedom.
Here, scalar, vector and tensor are those with respect to the three
dimensional isometry.
However, not all of them are physical. The fictitious gauge degrees of
freedom can be removed by imposing appropriate gauge conditions.
Here in this note we choose the flat gauge as we will explain
immediately below, neglecting the vector and tensor degrees of freedom.
 Their contributions, especially those of tensor perturbations, to the higher order correlation functions
 of the curvature perturbation enter only through loop corrections, which are highly suppressed.

At the beginning, we have $n+4$ scalar variables: $n$ from $n$
scalar field components, 4 from the metric. Since there are 1 temporal
and 1 spatial gauge transformations in the scalar sector, we
can eliminate 2 of them.
In the flat gauge,
we impose the conditions that the perturbations of three dimensional
spatial metric on each time slice vanish. The remaining
metric degrees of freedom are perturbations of the lapse
function and the shift vector. We denote them by $\xi^\alpha$
symbolically.
Further, by solving 2 constraint equations, we
can also remove the remaining two degrees of freedom $\xi^\alpha$,
so that after all $n$ degrees of freedom are left.
Namely, we can write all the metric degrees of
freedom solely in terms of the field fluctuations $\delta\phi^I$.

First let us formally expand the metric fluctuations
$\delta\xi^\alpha$ in $\epsilon$ as
\begin{equation}
\xi^\alpha(\lambda=\epsilon) =
   \xi^\alpha_{0}+ \xi_{(1)}^\alpha\, \epsilon
+ \xi_{(2)}^\alpha\, \epsilon^2+\cdots\, .
\end{equation}
The constraint equations are simply given by the variation of the action
with respect to $\xi^\alpha$,
\begin{equation}
\label{constraint_eq}
\frac{\delta{S}}{\delta\xi^\alpha} = 0 \, .
\end{equation}
When we expand the action with respect to $\xi_{(n)}^\alpha$,
the $n$-th
order term in $\xi^\alpha$, we find
\begin{equation}
S=S|_{\xi_{(n)}^\alpha=0}
+\left.{\delta S\over\delta
  \xi_{(n)}^\alpha}\right\vert_{\xi_{(n)}^\mu =0} \!\!\!
   \xi_{(n)}^\alpha
+{1\over 2}\left.{\delta^2 S\over\delta
 \xi_{(n)}^\alpha \delta \xi_{(n)}^\beta}
  \right\vert_{\xi_{(n)}^\mu=0} \!\!\!
   \xi_{(n)}^\alpha
   \xi_{(n)}^\beta+\cdots.
\label{Sxin}
\end{equation}
Then, writing (\ref{constraint_eq}) as
\begin{equation}
\left.\frac{\delta{S}}{\delta\xi^\alpha}\right\vert_{\xi_{(n)}^\mu=0}
 = -\left.\frac{\delta^2{S}}{\delta\xi^\beta
     \delta\xi^\alpha}\right\vert_{\xi_{(n)}^\mu= 0}
     \!\!\!\! \xi^\beta_{(n)}
-{1\over 2}\left.\frac{\delta^3{S}}{\delta\xi^\beta
     \delta\xi^\gamma \delta\xi^\alpha}\right\vert_{\xi_{(n)}^\mu= 0}
     \!\!\!\! \xi^\beta_{(n)} \xi^\gamma_{(n)}+\cdots
=\mathcal{O}(\epsilon^n)  \, ,
\end{equation}
we find that both the second and the third terms on the right hand side
of (\ref{Sxin}) are $\mathcal{O}(\epsilon^{2n})$.
Hence, when we want to know the action to, say, the cubic order in $\epsilon$,
the second and higher order of $\xi^\alpha$ are not necessary.
To obtain the linear order of $\xi^\alpha$,
we only need to solve
the constraint equations (\ref{constraint_eq}) expanded up to linear
order in $\epsilon$,
\begin{equation}
\left(
\frac{\delta{S}}{\delta\xi^\alpha}\right)_{(1)} = 0 \, .
\end{equation}
Plugging the solution for $\xi_{(1)}^\alpha$
of the above constraint equations back into the action, we obtain
the action written in terms of the field perturbation $Q^I$.

\subsection{Explicit calculations}

Now let us move onto more explicit computations. We consider
a general matter Lagrangian which describes multi-field system
minimally coupled to Einstein gravity in the Arnowitt-Deser-Misner form~\cite{Arnowitt:1962hi},
\begin{equation}
S = \int d^4x N\sqrt{\gamma} \left\{ \frac{\mpl^2}{2} \left[ R^{(3)} + \frac{1}{N^2} \left( E^i{}_jE^j{}_i - E^2 \right) \right] + P \right\} \, ,
\end{equation}
where $R^{(3)}$ is the 3-curvature scalar constructed from the spatial metric $\gamma_{ij}$, and
\begin{equation}
E_{ij} \equiv \frac{1}{2} \left( \dot\gamma_{ij} - N_{i|j} - N_{j|i} \right) \, ,
\end{equation}
with a vertical bar denoting a covariant differentiation
with respect to $\gamma_{ij}$.
The gauge we choose is, as advertised,
the so-called flat gauge, in which the spatial metric
$\gamma_{ij}$ is unperturbed, i.e.
\begin{equation}
\gamma_{ij} = a^2\delta_{ij} \, ,
\end{equation}
which completely fixes both spatial slicing and temporal threading
beyond linear level~\cite{Noh:2004bc}, as long as one neglects the vector and tensor
perturbations.
In this gauge, we separate the action into the gravity and matter sectors,
\begin{equation}
\label{action_explicit}
S = S^{\rm (G)}+S^{\rm (M)},
\end{equation}
with
\begin{align}
S^{\rm (G)}=& \int d^4x\, {a^3\mpl^2 \over 2N} \left( E^i{}_jE^j{}_i - E^2 \right)
\nonumber\\
= & \int d^4x {a^3 \mpl^2\over N} \left[ -3H^2 +
2H N^i{}_{,i} + {1\over 4} \left(
N_{i,j}N^{i,j} + N_{i,j}N^{j,i} - 2N^i{}_{,i}N^j{}_{,j} \right)\right]
\, ,\label{gravaction}
\\
S^{\rm (M)}= &\int d^4 x\, a^3 N P\, .
\end{align}
We choose
the background values of the metric variables, which we associate with a
subscript (0), as $N_{(0)} = 1$ and $N_{(0)}^i = 0$, corresponding to the
Friedmann-Lema\^itre-Robertson-Walker
model written using the cosmological time coordinate.
As we have explained above, to obtain the cubic order action,
we only need to keep the linear order for the metric perturbations $\xi^\alpha$.
In the action (\ref{action_explicit}), therefore we set
\begin{align}
N  = & 1 + N_{(1)}\epsilon \, ,
\label{xifirst1}
\\
N^i = & N^i_{(1)} \epsilon \, .
\label{xifirst2}
\end{align}

\subsubsection{Action expansion including metric perturbations}

It is straightforward to write down
the gravity part of the action.
All we need to do is just to plug the expansions (\ref{xifirst1}) and (\ref{xifirst2}) into
the gravity action (\ref{gravaction}). To the cubic order, we have
\begin{align}
S^{\rm (G)}=&  \int d^4x\,
a^3 \mpl^2 \left[
1-N_{(1)}\epsilon +N_{(1)}^2\epsilon^2-N_{(1)}^3\epsilon^3
\right]
\cr
&\times \left[ -3H^2 +
2H N^{i}_{(1),i}\, \epsilon + \frac{1}{4} \left(
N^{(1)}_{i,j}N_{(1)}^{i,j} + N^{(1)}_{i,j}N_{(1)}^{j,i} - 2N^{i}_{(1),i}N^{j}_{(1),j} \right)\epsilon^2\right]
\, .
\end{align}

The expansion of the matter Lagrangian is a little more non-trivial.
However, by assumption, our matter Lagrangian contains the spacetime metric
only through $X^{IJ}$. Therefore, all we have to do is just to replace
the expression for $\delta X^{IJ}$ to the one that explicitly includes
the expansion with respect to metric perturbations.
As the inverse metric $g^{\mu\nu}$ is given by
\begin{equation}
 g^{\mu\nu}\partial_\mu\partial_\nu=-{1\over
  N^2}(\partial_t-N^j\partial_j)^2+\gamma^{ij}\partial_i\partial_j \, ,
\end{equation}
(\ref{expansionX1}), (\ref{expansionX2}) and (\ref{expansionX3}) are more explicitly written down as
\begin{align}
D_\lambda X^{IJ}|_{\lambda=0} = &
 {1\over N^2} \widetilde D_tQ^{(I}\dot\phi_0^{J)}
\, ,
\\
D_\lambda^2X^{IJ}|_{\lambda=0} = &
{1\over N^2}\left[
R^{(I}{}_{KLM} \dot\phi_0^{J)}\dot\phi_0^MQ^KQ^L + \widetilde D_tQ^I \widetilde
 D_tQ^J\right]  -
 \gamma^{ij}\partial_iQ^I\partial_jQ^J
\, ,
\\
D_\lambda^3X^{IJ}|_{\lambda=0} = &
 {1\over N^2}\bigl[
R^{(I}{}_{KLM;N}\dot\phi_0^{J)}\dot\phi_0^MQ^NQ^KQ^L +
 R^{(I}{}_{KLM}\dot\phi_0^{J)}Q^KQ^L \widetilde D_tQ^M
\cr
&\qquad +
 3R^{(I}{}_{KLM}\dot\phi_0^MQ^KQ^L\widetilde D_tQ^{J)}
\bigr]\, ,
\end{align}
where we have defined
\begin{equation}
\widetilde D_t \equiv D_t-N^j\partial_j \, .
\end{equation}
More explicitly expanding the perturbation of $X^{IJ}$
in terms of $\epsilon$, we obtain
\begin{equation}
X^{IJ}=X^{IJ}_0+X^{IJ}_{(1)}\epsilon
 +X^{IJ}_{(2)}\epsilon^2 +X^{IJ}_{(3)}\epsilon^3+\cdots,
\end{equation}
with
\begin{align}
 X^{IJ}_{(1)}= &
 -N_{(1)}\dot\phi_0^I\dot\phi_0^J+D_t Q^{(I}\dot\phi{}_0^{J)} \, ,
 \\
 X^{IJ}_{(2)}= &
 {3\over 2}N_{(1)}^2 \dot\phi_0^I\dot\phi_0^J
  - 2 N_{(1)} D_t Q^{(I}\dot\phi{}_0^{J)}
  - N^j_{(1)} \partial_j Q^{(I}\dot\phi{}_0^{J)}
  \nonumber\\
   & +{1\over 2}\left[
  D_t Q^{(I} D_t Q^{J)}-\gamma^{ij}\partial_i Q^{(I}\partial_j Q^{J)}
   +R^{(I}_{~KLM}\dot\phi{}_0^{J)}\dot\phi_0^M Q^K Q^L
    \right] \, ,
 \\
 X^{IJ}_{(3)}= &
 -2 N_{(1)}^3 \dot\phi_0^I\dot\phi_0^J
  + 3 N_{(1)}^2 D_t Q^{(I}\dot\phi{}_0^{J)}
   -N_{(1)}^j\partial_jQ^{(I} D_t Q^{J)}
  \nonumber\\
  & - N_{(1)}\left[
    D_t Q^{(I} D_t Q^{J)}
   +R^{(I}_{~KLM}\dot\phi{}_0^{J)}\dot\phi_0^M Q^K Q^L
     - 2 N^j_{(1)} \partial_j Q^{(I}\dot\phi{}_0^{J)}
   \right]
   \nonumber\\
&+{1\over 6}\left[
   3R^{(I}_{~KLM}D_t Q^{J)}\dot\phi_0^M Q^K Q^L
  +R^{(I}_{~KLM}\dot\phi{}_0^{J)}D_t Q^M Q^K Q^L
  +R^{(I}_{~KLM;N}\dot\phi{}_0^{J)}\dot\phi_0^M Q^N Q^K Q^L
\right]\, .
\end{align}

\subsubsection{Linear order action}

First we consider the first order terms, where we can extract the
background equations of motion. Collecting the
results that we have obtained in the preceding sections, the first
order action becomes
\begin{equation}\label{S1}
S_1 = \int d^4x a^3 \left[ \left( 3\mpl^2H^2 + P_0 - P_{\langle{IJ}\rangle}\dot\phi_0^I\dot\phi_0^J \right) N_{(1)} + P_{\langle{IJ}\rangle}D_tQ^I\dot\phi_0^J + P_af_{a;I}Q^I \right] \, ,
\end{equation}
where by $P_0$ we denote the matter Lagrangian with the background
quantities substituted. What we can immediately see is
that we can derive two equations of motion by varying with respect to the
lapse perturbation $N_{(1)}$ and the field fluctuation
$Q^I$. Taking a variation of (\ref{S1}) with respect to $N_{(1)}$,
we obtain
\begin{equation}
H^2 = \frac{1}{3\mpl^2} \left( P_{\langle{IJ}\rangle}\dot\phi_0^I\dot\phi_0^J - P_0 \right) \, ,
\end{equation}
which is the background Friedmann equation. We can also immediately obtain the equation of the background field $\phi_0^I$ as
\begin{equation}
\frac{1}{a^3}D_t \left( a^3P_{\langle{IJ}\rangle}\dot\phi_0^J \right) = P_af_{a;I} \, ,
\end{equation}
or more explicitly,
\begin{equation}
\left( P_{\langle{IJ}\rangle} + P_{\langle{IK}\rangle\langle{JL}\rangle}\dot\phi_0^K\dot\phi_0^L \right) D_t\dot\phi_0^J + \left( 3HP_{\langle{IJ}\rangle} + P_{\langle{IJ}\rangle{a}}f_{a;K}\dot\phi_0^K \right) \dot\phi_0^J - P_af_{a;I} = 0 \, .
\end{equation}

\subsubsection{Quadratic order action}

Again after straightforward
manipulations, we find
\begin{align}\label{S2}
S_2 = & \int d^4x\, a^3 \bigg\{ \frac{1}{2} \bigg[ P_{\langle{IJ}\rangle}
 \left( R^I{}_{KLM}\dot\phi_0^J\dot\phi_0^MQ^KQ^L + D_tQ^ID_tQ^J -
 \gamma^{ij}\partial_iQ^I\partial_jQ^J \right) + P_af_{a;IJ}Q^IQ^J
\nonumber\\
& \hspace{2.3cm} +
 P_{\langle{IJ}\rangle\langle{KL}\rangle}D_tQ^I\dot\phi_0^JD_tQ^K\dot\phi_0^L + 2P_{\langle{IJ}\rangle{a}}D_tQ^I\dot\phi_0^Jf_{a;K}Q^K + P_{ab}f_{a;I}f_{b;J}Q^IQ^J \bigg]
\nonumber\\
& \hspace{1.8cm} + N_{(1)}\left[ -P_{\langle{IJ}\rangle}D_tQ^I\dot\phi_0^J +
 P_af_{a;I}Q^I - \left(
 P_{\langle{IJ}\rangle\langle{KL}\rangle}D_tQ^K\dot\phi_0^L +
 P_{\langle{IJ}\rangle{a}}f_{a;K}Q^K \right) \dot\phi_0^I\dot\phi_0^J \right]
\nonumber\\
& \hspace{1.8cm} + \frac{N_{(1)}^2}{2}
 \left( -6\mpl^2H^2 +
 P_{\langle{IJ}\rangle}\dot\phi_0^I\dot\phi_0^J +
 P_{\langle{IJ}\rangle\langle{KL}\rangle}\dot\phi_0^I
 \dot\phi_0^J\dot\phi_0^K\dot\phi_0^L \right)
  - 2\mpl^2HN_{(1)}N_{(1),i}^{i}
\nonumber\\
& \hspace{1.8cm} -
 P_{\langle{IJ}\rangle}N_{(1)}^i \partial_iQ^I\dot\phi_0^J
 + \frac{\mpl^2}{4} \left( N^{(1)}_{i,j} N_{(1)}^{i,j} +
 N^{(1)}_{i,j} N_{(1)}^{j,i} - 2N_{(1),i}^{i} N_{(1),j}^{j}
 \right) \bigg\} \, .
\end{align}

From the second order action (\ref{S2}), we can derive the linear order
metric perturbations, $N_{(1)}$ and $N_{(1)}^i$, which we have not
specified yet.
Varying the quadratic action (\ref{S2})
with respect to $N_{(1)}^i$ and $N_{(1)}$, we obtain
the constraint equations which is easily solved as
\begin{align}
\label{lapse_sol}
N_{(1)} = & \frac{1}{2\mpl^2H}P_{\langle{IJ}\rangle}Q^I\dot\phi_0^J \equiv \calN_IQ^I  \, ,
\\
-2\mpl^2H\frac{\Delta}{a^2}\chi = & N_{(1)} \left(
 P_{\langle{IJ}\rangle}\dot\phi_0^I\dot\phi_0^J - 2P_0 -
 P_{\langle{IJ}\rangle\langle{KL}\rangle}\dot\phi_0^I\dot\phi_0^J\dot\phi_0^K\dot\phi_0^L
 \right)
\nonumber\\
& + \left( P_{\langle{IJ}\rangle} + P_{\langle{IJ}\rangle\langle{KL}\rangle}\dot\phi_0^K\dot\phi_0^L \right)D_tQ^I\dot\phi_0^J + \left( -P_af_{a;I}+P_{\langle{JK}\rangle{a}}f_{a;I}\dot\phi_0^J\dot\phi_0^K \right)Q^I \, ,
\end{align}
where we have set $N_{(1)i} = \partial_i\chi$, which is allowed when
we consider only scalar perturbations.

\subsubsection{Cubic order action}

Now we turn to the third order action.
First, we present the contributions coming from the gravity sector
$S^{\rm (G)}$.
We can easily collect the third order terms to obtain
\begin{equation}
S_3^\text{(G)} = \int d^4x a^3 \bigg\{ 3\mpl^2H^2  N_{(1)}^3 + 2\mpl^2H
\frac{\Delta}{a^2}\chi N_{(1)}^2 -
\frac{\mpl^2}{2a^4}\left[ \chi^{,ij}\chi_{,ij} - (\Delta\chi)^2 \right]
N_{(1)} \bigg\} \, .
\end{equation}
Next we consider the contributions from matter sector, $S^{\rm (M)}$.
Using the notation introduced in (\ref{lapse_sol}),
after some arrangement we find
\begin{align}
S_3^\text{(M)} = & \int d^4x a^3 \Big[ (g_1)_{IJK}Q^IQ^JQ^K + (g_2)_{IJK}D_tQ^IQ^JQ^K + (g_3)_{IJK}D_tQ^ID_tQ^JQ^K
\nonumber\\
& \hspace{1.6cm} + (g_4)_{IJK}D_tQ^ID_tQ^JD_tQ^K + (g_a)_{IJ}Q^I\partial_iQ^JN_{(1)}^i + (g_b)_{IJ}D_tQ^I\partial_iQ^JN_{(1)}^i
\nonumber\\
& \hspace{1.6cm} + (g_c)_{IJK}Q^I\gamma^{ij}\partial_iQ^J\partial_jQ^K + (g_d)_{IJK}D_tQ^I\gamma^{ij}\partial_iQ^J\partial_jQ^K \Big] \, ,
\end{align}
where
\begin{align}
(g_1)_{IJK} = & \frac{1}{6} \Big( P_{\langle{LM}\rangle}R^L{}_{IJN;K}\dot\phi_0^M\dot\phi_0^N + P_af_{a;IJK} + 3P_{\langle{LM}\rangle{a}}R^L{}_{IJN}\dot\phi_0^M\dot\phi_0^Nf_{a;K}
\nonumber\\
& \hspace{0.5cm} + 3P_{ab}f_{a;IJ}f_{b;K} + P_{abc}f_{a;I}f_{b;J}f_{c;K} \Big)
\nonumber\\
& + \frac{1}{2}\calN_K \Big[ -P_{\langle{LM}\rangle}R^L{}_{IJN}\dot\phi_0^M\dot\phi_0^N + P_af_{a;IJ} + P_{ab}f_{a;I}f_{b;J}
\nonumber\\
& \hspace{1.5cm} - \dot\phi_0^L\dot\phi_0^M \left( P_{\langle{LM}\rangle\langle{AB}\rangle}R^A{}_{IJC}\dot\phi_0^B\dot\phi_0^C + P_{\langle{LM}\rangle{a}}f_{a;IJ} + P_{\langle{LM}\rangle{ab}}f_{a;I}f_{b;J} \right) \Big]
\nonumber\\
& + \frac{1}{2}\calN_J\calN_K \left( P_{\langle{AB}\rangle{a}}f_{a;I}\dot\phi_0^A\dot\phi_0^B + P_{\langle{AB}\rangle\langle{CD}\rangle{a}}f_{a;I}\dot\phi_0^A\dot\phi_0^B\dot\phi_0^C\dot\phi_0^D \right)
\nonumber\\
& - \calN_I\calN_J\calN_K \left( \frac{1}{2}P_{\langle{AB}\rangle}\dot\phi_0^A\dot\phi_0^B + P_{\langle{AB}\rangle\langle{CD}\rangle}\dot\phi_0^A\dot\phi_0^B\dot\phi_0^C\dot\phi_0^D + \frac{1}{6} P_{\langle{AB}\rangle\langle{CD}\rangle\langle{EF}\rangle} \dot\phi_0^A\dot\phi_0^B\dot\phi_0^C\dot\phi_0^D\dot\phi_0^E\dot\phi_0^F \right) \, ,
\\
(g_2)_{IJK} = & \frac{1}{6} \left( P_{\langle{LM}\rangle}R^L{}_{JKI} + 3P_{\langle{IL}\rangle}R^L{}_{JKM} \right) \dot\phi_0^M + \frac{1}{2}P_{\langle{IL}\rangle{a}}\dot\phi_0^Lf_{a;JK} + \frac{1}{2}P_{\langle{IL}\rangle\langle{AB}\rangle}R^A{}_{JKM}\dot\phi_0^B\dot\phi_0^L\dot\phi_0^M
\nonumber\\
& + \frac{1}{2}P_{\langle{IL}\rangle{ab}}\dot\phi_0^Lf_{a;J}f_{b;K} - \calN_K \left( P_{\langle{IL}\rangle{a}}\dot\phi_0^Lf_{a;J} + P_{\langle{IL}\rangle\langle{MN}\rangle{a}}\dot\phi_0^L\dot\phi_0^M\dot\phi_0^Nf_{a;J} \right)
\nonumber\\
& + \calN_J\calN_K \left( P_{\langle{IL}\rangle}\dot\phi_0^L + \frac{5}{2}P_{\langle{IL}\rangle\langle{MN}\rangle}\dot\phi_0^L\dot\phi_0^M\dot\phi_0^N + \frac{1}{2}P_{\langle{IL}\rangle\langle{MN}\rangle\langle{AB}\rangle} \dot\phi_0^L\dot\phi_0^M\dot\phi_0^N\dot\phi_0^A\dot\phi_0^B \right) \, ,
\\
(g_3)_{IJK} = & -\frac{1}{2}\calN_K \left[ P_{\langle{IJ}\rangle} + \left( P_{\langle{IJ}\rangle\langle{LM}\rangle} + 3P_{\langle{IL}\rangle\langle{JM}\rangle} \right) \dot\phi_0^L\dot\phi_0^M + P_{\langle{IL}\rangle\langle{JM}\rangle\langle{AB}\rangle} \dot\phi_0^L\dot\phi_0^M\dot\phi_0^A\dot\phi_0^B \right]
\nonumber\\
& + \frac{1}{2} \left( P_{\langle{IJ}\rangle{a}} + P_{\langle{IL}\rangle\langle{JM}\rangle{a}} \dot\phi_0^L\dot\phi_0^M \right) f_{a;K} \, ,
\\
(g_4)_{IJK} = & \frac{1}{2}P_{\langle{IJ}\rangle\langle{KL}\rangle}\dot\phi_0^L + \frac{1}{6}P_{\langle{IL}\rangle\langle{JM}\rangle\langle{KN}\rangle}\dot\phi_0^L\dot\phi_0^M\dot\phi_0^N \, ,
\\
(g_a)_{IJ} = & \calN_I \left( P_{\langle{JK}\rangle}\dot\phi_0^K + P_{\langle{JK}\rangle\langle{LM}\rangle}\dot\phi_0^K\dot\phi_0^L\dot\phi_0^M \right) - P_{\langle{JK}\rangle{a}}f_{a;I}\dot\phi_0^K \, ,
\\
(g_b)_{IJ} = & -P_{\langle{IJ}\rangle} - P_{\langle{IK}\rangle\langle{JL}\rangle}\dot\phi_0^K\dot\phi_0^L \, ,
\\
(g_c)_{IJK} = & \frac{1}{2}\calN_I \left( -P_{\langle{JK}\rangle} + P_{\langle{JK}\rangle\langle{LM}\rangle}\dot\phi_0^L\dot\phi_0^M \right) - \frac{1}{2}P_{\langle{JK}\rangle{a}}f_{a;I} \, ,
\\
(g_d)_{IJK} = & -\frac{1}{2}P_{\langle{IL}\rangle\langle{JK}\rangle}\dot\phi_0^L \, .
\end{align}

\subsection{Effects of field space geometry}

As we have computed up to the cubic order action in the covariant form,
now we can easily appreciate the effects of field space geometry. An
elementary consideration comes from the second order action
(\ref{S2}). For this purpose, we may restrict ourselves to the simplest
case of a canonical two-field model where the matter Lagrangian is given
by $P=G_{IJ}X^{IJ}-V$ with a constant field space curvature $R^I{}_{JKL}
= K \left( \delta^I{}_K G_{JL} - \delta^I{}_L G_{JK} \right)$, with $K$
being a constant called Gaussian curvature: this form of the curvature
tensor describes a two-dimensional surface with a constant
curvature. Further, we can choose the basis in such a way that one is
pointing along and the other is orthogonal to the field trajectory, so
that we may interpret the former as the curvature mode $\sigma$ and the
latter the isocurvature mode $s$ with $G_{IJ}$ being diagonal,
i.e. $G_{\sigma{s}}=0$. Then, the curvature term in (\ref{S2}) becomes
\begin{equation}\label{example}
S_2 \supset \int d^4x \frac{a^3}{2} R_{IKLJ}\dot\phi_0^K\dot\phi_0^LQ^IQ^J \, ,
\end{equation}
with the field space index either $\sigma$ or $s$. By the symmetry of the Riemann curvature tensor, the only non-zero component is $R_{\sigma{s}\sigma{s}}$. Further, by definition $\dot\phi_0^s=0$, since the isocurvature mode remains always orthogonal to the trajectory. Thus, the only non-zero contribution is
\begin{eqnarray}
S_2 \supset
\int d^4x \frac{a^3}{2} R_{s\sigma\sigma s}
\dot\phi_0^\sigma\dot\phi_0^\sigma Q^sQ^s=
-\frac {K}{2} \int d^4x\, a^3 \dot\phi_0^2 \,(Q^s)^2 \, .
\end{eqnarray}
Thus, we can immediately see that only the perturbation in the
isocurvature mode is affected by the field space curvature, either
enhanced ($K<0$) or suppressed ($K>0$)
depending on the signature of the curvature,
while that in the curvature mode remains intact.
 Such a constant, negative curvature can be realized, for example, for the motion of a D-brane
 in the internal anti de Sitter space.

To generate significantly large contribution of isocurvature
perturbation at the epoch when the relevant scales cross the horizon
during inflation, the mass squared in this direction should be
suppressed compared with $H^2$. Otherwise, it decays exponentially.
It is, however, hard to imagine that
the mass squared is largely negative because the background trajectory will
be unstable. The region with negative
mass squared cannot extend indefinitely, and it should be
surrounded by the regions where the mass squared is positive.
To keep the trajectory along the region with the mass squared
negative, it is difficult to avoid the tuning problem of the
initial conditions for the background trajectory.
Therefore, it would be natural to assume that the mass squared is
non-negative at the early stage, and then the corresponding isocurvature
perturbation is not amplified during its super-horizon evolution.
At a later epoch, the mass squared in the initial isocurvature
direction can become negative.
However, if the mass squared is largely negative, the field rapidly
rolls away from the initial isocurvature direction.
Thus, the stage in which the mass squared is negative
will not last long. Therefore it is difficult to
selectively enhance the initial isocurvature perturbation
during inflation.

 It is, however, not impossible to enhance the isocurvature perturbation by incorporating
 the field space curvature as follows.
The typical size of the mass squared induced
by the field space curvature would be
\begin{equation}
m^2_{\rm eff}
\sim
\dot\phi^2R
\sim \epsilon \beta H^2 \, ,
\end{equation}
where $\epsilon$ is the standard slow-roll parameter and
$\beta \equiv R \mpl^{2}$ is the ratio of the field space curvature to
its typical value in the context of supergravity.
 Now, we consider the case of negative curvature. Then,
the
effective mass squared of the isocurvature perturbation may be negative
even if
the background trajectory keeps along the valley of the potential with the
``bare'' mass squared of the isocurvature perturbation positive.
In this case,
we can make the effective mass squared negative everywhere without fine
tuning of the background trajectory.
The magnitude of the magnification effect due to
this effective negative mass squared is evaluated by the integral
\begin{equation}
\exp \left( \int \frac{m^2_{\rm eff}}{H^2} dN \right) \sim e^{\int
 \epsilon\beta dN} \, .
\end{equation}
Using the estimate
$\epsilon\sim 1/\Delta N$ valid for the standard slow-roll,
where $\Delta N$ is the $e$-folding number during inflation,
the amplification effect is already marginally significant for $\beta=1$.
If we have a negative $R$ with the magnitude being larger
than $\mpl^{-2}$, the curvature effect can easily
give rise to large amplification of the isocurvature perturbation.

Before closing this section, we should also mention the effects by
curved background trajectories~\cite{curvedtraj}, which is another genuine
phenomenon in multi-field system. A convenient way of describing
perturbation around a
curved trajectory is to introduce the decomposition into curvature and
isocurvature modes, i.e. to construct a set of bases which is moving
with the trajectory, with one of them pointing along and
the others being orthogonal to the trajectory.
An advantage of using such decomposition
is its clear meaning throughout the evolution of perturbation.
However, even when all components of the mass matrix are negligible small,
in general, we have continuous mixing between curvature and
isocurvature modes when the trajectory is curved.

We can consider an alternative to the decomposition of the
curvature-isocurvature modes. It comes from the observation that in the
presence of the potential the equation of the background trajectory is
not the geodesic with respect to the field space metric. This makes
it impossible to introduce such a convenient coordinate system that
erases the Christoffel symbol along the trajectory, with one basis
vector being identical to the direction of the background
trajectory. Instead, we can introduce coordinates by parallelly
transporting the basis vectors chosen at an arbitrary time as
\begin{equation}\label{new_coord}
D_t e^I_a=0 \, ,
\end{equation}
where $a$ represents the new tetrad frame indices. In this case
Christoffel symbol does not vanish even on the background
trajectory. Nevertheless, it looks more convenient to use such
coordinates since the computation becomes more economical and
intuitive. The covariant derivatives acting on the perturbation variable
$Q^I$ all appear in the form of $D_t Q^I$. Using (\ref{new_coord}), we
find that
\begin{equation}
e_I^a D_t Q^I = \partial_t Q^a \, ,
\end{equation}
where $Q^a \equiv e_I^a Q^I$.
Namely, those derivatives become ordinary partial derivatives. If we use
such coordinates, all the information about
the linear evolution of perturbation is confined in
the effective mass matrix projected onto this tetrad frame.
Since the effective mass matrix is not diagonal in general,
the calculation is not so straightforward.
But still the description in this manner will help
our intuitive understanding of the effect of curved trajectories
in curved field space.
 It may deserve further study, which is beyond the scope of the present paper.

\section{Conclusions}
\label{sec:conclusions}

In this note, we have studied a covariant formulation of general multi-field inflation. Starting from the geodesic equation parametrized by $\lambda$ which connects a point on the background trajectory to the corresponding point with field perturbations, we have found the non-linear relation between the real physical field fluctuation $\delta\phi^I$ and the vector $Q^I$ living on the tangent space. Using this relation, we have expanded the general matter Lagrangian $P(G_{IJ},X^{IJ},\phi^I)$ in terms of $\lambda$ up to cubic order in $Q^I$. The resulting expression is fully covariant with the Riemann curvature tensor $R_{IJKL}$ describing the geometry of the field space.

Including gravity, we have chosen the flat gauge where metric perturbations are given by the solutions of the constraint equations in terms of $Q^I$. For an explicit calculation up to cubic order, which is necessary to find the leading contribution to the bispectrum of the curvature perturbation, we need only the linear solutions of the metric perturbation which could be found from the second order action. With these solutions, we have explicitly computed the cubic order action in a fully covariant manner. Although we have presented up to cubic order action, our formulation can be straightforwardly extended to find arbitrary higher order action. We have also discussed briefly the genuine effects in multi-field inflation generated by the isocurvature perturbations.

\subsection*{Acknowledgement}

JG thanks Ana Ach\'ucarro for important conversations.
JG is grateful to the Yukawa Institute for Theoretical Physics at Kyoto University for hospitality during  the long-term workshop ``Gravity and Cosmology 2010 (GC2010)'' (YITP-T-10-01) and the YKIS symposium ``Cosmology -- The Next Generation --'' (YKIS2010), where this work was initiated, and the 20th Workshop on General Relativity and Gravitation in Japan (YITP-W-10-10) where this work was under progress.
This work was supported in part by a Korean-CERN fellowship,
the Japanese Society for Promotion of Science Grants N. 21244033, the Global COE
Program ``The Next Generation of Physics, Spun from Universality and Emergence'',
and the Grant-in-Aid for Scientific Research on Innovative Areas (Ns. 21111006 and 22111507) from
the MEXT.

\appendix

\renewcommand{\theequation}{\thesection.\arabic{equation}}
\setcounter{equation}{0}

\section{Comparison with non-covariant expression}

Here we show a method to derive our new covariant expression from the
previously known non-covariant expression~\cite{Langlois:2008qf},
in order to clarify the equivalence between them.
One trivial replacement is to change all the partial differentiations
with respect to $\phi^I$ to the corresponding covariant ones.
A non-trivial point is that the covariant
expression contains the terms depending on the curvature of the field
space, $R^A{}_{BCD}$.

To obtain the terms with curvature,
we focus on the fact that $R^I{}_{ABJ}Q^A Q^B$
contains a term with second derivative of the metric contracted
with $Q^A$ in the form $G_{KJ,AB}Q^A Q^B$,
\begin{equation}
R^I{}_{ABJ}Q^A Q^B\supset
{1\over 2}G^{IK}G_{KJ,AB}Q^A Q^B\,.
\end{equation}
Such second derivatives of the field space metric arise in
the non-covariant expression from the second or higher
derivatives of $P$.
In the covariant formulation, differentiations acting on
the field space metric vanish by definition, while
they do not in the non-covariant notation.
Here, we recall that we are assuming that $P$ is a function of
$X^I{}_J \equiv X^{IK}G_{KJ}$ and scalar functions $f_a(\phi^A)$.
Indices among $X^I{}_J$ should be completely contracted in $P$,
i.e. there is no other quantity having the field space indices in $P$.
Using this fact, derivatives of $P$
with respect to $G_{KJ}$ can be related to those with respect to
$X^{IK}$. Namely, we have
\begin{equation}
 P_{,A}\supset {\partial P\over \partial X^I{}_J}X^{IK}G_{KJ,A}
 = G^{LJ} P_{\langle IL\rangle} X^{IK}G_{KJ,A} \, .
\end{equation}
Therefore, we find that $P_{,AB}$ contains a part of
curvature contribution,
\begin{equation}
 P_{,AB}\supset G^{LJ} P_{\langle LI\rangle} X^{IK}G_{KJ,AB}
\approx 2P_{\langle LI\rangle} X^{LK}R^I{}_{ABK}\,.
\end{equation}
Here $A$ and $B$ indices are understood to be contracted with $Q^A$ and $Q^B$, and
``$\approx$'' means the equality that is valid focusing only on
the term $G_{KJ,AB}Q^A Q^B$, neglecting the other terms in the
curvature.
In a similar way, we have
\begin{align}
 P_{,ABC} \supset & G^{LJ} P_{\langle LI\rangle} X^{IK}G_{KJ,ABC}
    +\left[G^{LJ} P_{\langle LI\rangle,C} X^{IK}G_{KJ,AB}
        +\Big(2\mbox{ permutations among } A,B,C\Big)\right]\cr
    \approx &
 2P_{\langle LI\rangle} X^{LK}R^I{}_{ABK;C}
+ \left[2P_{\langle LI\rangle;C} X^{LK}R^I{}_{ABK}
        +\Big(2\mbox{ permutations among } A,B,C\Big)\right]\,,
\end{align}
and
\begin{align}
 P_{\langle IJ\rangle,AB} \supset &
{\partial\over\partial X^{IJ}}\left(
   G^{NL} P_{\langle LM\rangle} X^{MK}G_{KN,AB}\right)\cr
=& G^{NL} P_{\langle IJ\rangle \langle LM\rangle} X^{MK}G_{KN,AB}
 +{1\over 2}\left(
  G^{KL} P_{\langle IL\rangle} G_{KJ,AB}
  +  G^{KL} P_{\langle JL\rangle} G_{KI,AB}\right)\cr
    \approx &
 2P_{\langle IJ\rangle\langle LM\rangle} X^{MK}R^L{}_{ABK}+
 P_{\langle IL\rangle} R^L{}_{ABJ}+
 P_{\langle JL\rangle} R^L{}_{ABI}\,.
\end{align}

There is another origin of $G_{KJ,AB}Q^A Q^B$.
The field perturbation introduced in non-covariant formulation
$\delta\phi^A$ is related to our $Q^A$ by (\ref{mapping2}),
\begin{equation}
\delta\phi^A=Q^A-{1\over 2}\Gamma^A_{\,IJ}Q^I Q^J-{1\over 6}G^{AI}
    G_{IJ,KL}Q^J Q^K Q^L+\cdots,
\end{equation}
where we have abbreviated several terms at the cubic order,
except for the term containing the combination $G_{KJ,AB}Q^A Q^B$.
One may think that this cubic order contribution is higher order
in action since the
perturbed action starts with the second order of perturbation.
However, the absence of linear terms in the perturbed action is
achieved only after using the background equation of motion.
The use of background equation of motion erases linear terms
in respective formulations, but the meaning
of linear terms varies in different formulations.
Therefore,
the perturbed actions in different formulation naturally
differ by the terms proportional to the background
equation of motion.
This explains that we have to take into account the linear
term in the non-canonical expression to obtain the
correct curvature correction. Discriminating the quantities
in the non-canonical formulation by associating an underbar, we have
\begin{align}
{\underbar P}_{(1)}
\supset&
P_{\langle IJ\rangle}
{\underbar X}{}_{(1)}^{IJ} \supset
P_{\langle IJ\rangle}\dot\phi^{(I}\delta\dot \phi^{J)}\cr
\supset&
 -{1\over 6} P_{\langle IJ\rangle}\dot\phi^I
  G^{JK}G_{KB,CD}\dot Q^B Q^C Q^D\cr
\approx&
 -{1\over 3} P_{\langle IJ\rangle}\dot\phi^I
  R^J{}_{CDB}\dot Q^B Q^C Q^D\,.
\end{align}
Following the rules mentioned above, all the terms in
the perturbed action with the field space curvature
can be reproduced correctly.

\end{document}